\documentclass[sigconf]{acmart}
\pdfoutput=1
\usepackage{booktabs}
\usepackage{graphicx}
\usepackage{algorithmic}
\usepackage{algorithm}
\usepackage{float}
\usepackage{url}

\copyrightyear{2017}
\acmYear{2017} 
\acmConference[XXX]{XXX}{XXX}{XXX}

\DeclareGraphicsExtensions{.pdf,.jpeg,.png}

\begin{document}
\title{Beyond 16GB: Out-of-Core Stencil Computations}

\author{Ist\'an Z Reguly}
\affiliation{
\institution{Faculty of Information Technology and Bionics,P\'azm\'any P\'eter Catholic University}
       \city{Budapest}
       \country{Hungary}}
\email{reguly.istvan@itk.ppke.hu}

\author{Gihan~R.~Mudalige}
\affiliation{
       \institution{Department of Computer Science, University of Warwick}
       \city{Coventry}
       \country{UK}}
       \email{g.mudalige@warwick.ac.uk}

\author{Michael~B.~Giles}
\affiliation{
       \institution{Maths Institute, University of Oxford}
       \city{Oxford}
       \country{UK}}
       \email{mike.giles@maths.ox.ac.uk}

\renewcommand{\shortauthors}{I. Z. Reguly et al.}

\begin{abstract}
Stencil computations are a key class of applications, widely used in the scientific computing community, and a class that has particularly benefited from performance improvements on architectures with high memory bandwidth. Unfortunately, such architectures come with a limited amount of fast memory, which is limiting the size of the problems that can be efficiently solved. In this paper, we address this challenge by applying the well-known cache-blocking tiling technique to large scale stencil codes implemented using the OPS domain specific language, such as CloverLeaf 2D, CloverLeaf 3D, and OpenSBLI. We introduce a number of techniques and optimisations to help manage data resident in fast memory, and minimise data movement. Evaluating our work on Intel's Knights Landing Platform as well as NVIDIA P100 GPUs, we demonstrate that it is possible to solve 3 times larger problems than the on-chip memory size with at most 15\% loss in efficiency.
\end{abstract}
%
%
%
%
%

\maketitle

\section{Introduction}
Today's accelerators offer unparalleled computational throughput, as well as high amounts of bandwidth to a limited amount of on-chip or on-board memory. The size of this fast memory has been a significant limiting factor in their adoption, as for most problem classes, it sets an upper bound for the problem sizes that can be solved on any single device. For larger problems, one had to either use multiple GPUs or fall back to the CPU, which usually has at least an order of magnitude larger memory.

Another significant limiting factor is the speed at which data can be uploaded to the accelerator memory. There is a great disparity between the bandwidth from a large memory to the accelerator memory (typically from CPU memory through PCI-e) and the bandwidth of the accelerator (up to 45$\times$). This traditionally meant that all data was uploaded to the accelerator memory initially, and stayed resident for the entirety of the application - yielding the aforementioned size limitation. 

In data streaming type applications, where a chunk of data is uploaded, processed, then downloaded, the workload (usually larger than GPU memory) is partitioned into small chunks, so it's possible to overlap copies in both directions and computations. To efficiently utilise accelerator bandwidth, this also means that any data uploaded has to be accessed about as many times as this ratio between upload bandwidth and accelerator bandwidth; otherwise performance will be limited by upload speed. To efficiently utilise the accelerator's computational resources, the ratio is even more extreme: for a P100 GPU one would need to carry out about 2500 floating point operations for every float variable uploaded (10 TFlops/s, 16 GB/s PCI-e BW, 4 bytes/float).

Going into the exascale era, most of the upcoming large supercomputers will be built with chips featuring on-chip high-bandwidth memory: Intel's Knight's Landing and later generations have at least 16GB MCDRAM, with bandwidths over 500 GB/s, and NVIDIA's P100 and later GPUs also feature 16GB with 720 GB/s and more. To tackle the issue with slow upload speeds, both have moved away from PCI-e: Intel's chips are stand-alone, have direct access to DDR4 memory (90GB/s on KNL), and the stacked memory can be used either as a separate memory space (flat mode) or as a large cache (cache mode). NVIDIA has introduced NVLink, connecting their GPUs to IBM CPUs and other GPUs, with 40 GB/s (in both directions), and allows oversubscribing GPU memory through Unified Memory - practically allowing stacked memory to become a large cache. While this significantly improves the upload/bandwidth ratio, and helps many applications, by no means does it solve the problem.

In this paper we present research that targets the memory size limitation challenge on structured mesh computations, a key class of applications mainly used for solving discretised partial differential applications. Our work is done in the framework of the OPS domain specific language (DSL) \cite{ops-github} embedded in C/Fortran, which presents a high-level abstraction for describing structured mesh algorithms, and automatically parallelises them for a range of parallel architectures using MPI, OpenMP, CUDA, OpenACC and OpenCL. OPS has been shown to deliver near-optimal performance, compared to carefully hand-coded versions \cite{ops1,ops2} of large-scale stencil codes, including CloverLeaf 2D, CloverLeaf 3D and OpenSBLI\cite{opensbli}.

Structured mesh stencil codes are generally bound by memory bandwidth not computational throughput, and on conventional CPU architectures loop tiling optimisations \cite{uday08cc,Tang:2011:PSC:1989493.1989508,SCHWEITZ2006RSTREAM,yask} have proven very effective in improving spatial and temporal locality, with the goal of improving cache utilisation. In previous work \cite{arxiv} we have shown that OPS can deploy such an optimisation at run-time even on large-scale codes, in contrast to existing compile-time tiling approaches which cannot cope with tiling across dynamic execution paths and multiple compilation units. To the best of our knowledge, cache-blocking tiling has not been evaluated in situations targeting stacked memory, and certainly not on applications the size of CloverLeaf or OpenSBLI.

In this paper, we make the following contributions:
\begin{enumerate}
	\item We adapt the cache-blocking tiling algorithm to target the stacked memories of the latest HPC architectures (KNL and P100).
	\item We evaluate explicit and implicit (unified memory) memory management strategies on PCI-e and NVLink.
	\item We carry out a problem scaling and performance analysis on the CloverLeaf 2D, 3D, and OpenSBLI codes.
\end{enumerate}

The rest of the paper is organised as follows: Section \ref{sec/related} discusses related work, Section \ref{sec/ops} introduces OPS, the lazy execution scheme and the dependency analysis used for cache-blocking tiling, Section \ref{sec/adapt} describes how tiling in OPS is targeting architectures with stacked memory, Section \ref{sec/perf} carries out the performance analysis, and Section \ref{sec/conclusions} draws conclusions.

\section{Related work}\label{sec/related}

There is already a considerable body of related work investigating performance on Intel's Knights Landing (KNL), and a select few evaluate performance in out-of-core scenarios. The work of Heinecke et. al. \cite{seismic} places some more frequently used datasets in MCDRAM and accesses less-frequently used ones in DDR4, achieving high efficiency - this is then compared to running in cache mode (where all of the MCDRAM is a large cache), which yields performance close to the explicit placement version. This demonstrates that for an application where datasets can be partitioned into frequently and less frequently used categories, Intel's strategy does lead to high memory bandwidth utilisation. Work by Vienne et. al. \cite{scalability} carries out a more detailed study of problem size scaling, showing that as long as the size is less than 16GB, there is very little difference between the cache and flat modes; LBS3D is 4.3$\times$ faster than running with DDR4 only, and miniFE is 3.1$\times$. However, as size grows beyond 16 GB, performance is falling off sharply: for LBS3D at 48 GB there is only a 1.19$\times$ speedup versus not using MCDRAM at all, and on miniFE at 28 GB only a speedup of 1.5$\times$. Work by Tobin et. al. \cite{Tobin2017} also evaluates scaling beyond 16GB with a seismic simulation code; compared to a 7GB dataset, running on a 20GB the GFlops achieved is reduced by a factor of 0.64$\times$, and at 39 GB it is reduced by .356$\times$ - while running with DDR4 only, the reduction is 0.21$\times$. Authors of \cite{berkeleygw, factorization} and many other papers focus on staying in the 16 GB MCDRAM to achieve high performance.

These issues are much more pronounced on GPUs, where the difference in upload bandwidth and on-device bandwidth are much larger (e.g. 720 GB/s device vs. 16 GB/s PCI-e or 40 GB/s NVLink 1.0 in the P100). Furthermore, GPU memory either has to be explicitly managed, or used in unified memory mode, where page faults on the GPU cause transfers of memory pages - this has much higher latency compared to cache misses on the KNL, and there is no automatic prefetch mechanism - though one can programmatically prefetch pages, improving efficiency. There are many examples of classical data streaming applications \cite{ska_stream} that work in the way described above. However, there are much fewer examples of trying to run out-of-core algorithms on GPUs - because of the PCI-e bottleneck. There are some computationally intensive algorithms where this is worth doing - such as the matrix-matrix multiplication, which serves as a basis for the work by Allombert et. al. \cite{ALLOMBERT2014888} which performs a tiled Cholesky factorisation - in this case there is enough computational work per byte uploaded. Similar streaming techniques are used for computationally intensive algorithms in \cite{Lin:2017qy}, and there are applications in visualisation as well \cite{Wang:2013:GOM:2508363.2508413,TGIS:TGIS1362}. With the introduction of Pascal, Unified Memory and GPU memory oversubscription, the work of Sakharnykh \cite{forall} demonstrates that on an Adaptive Mesh Refinement code, which is mostly bound by memory bandwidth, using careful annotations and prefetching it is possible to scale the problem size beyond 16GB. This required the fetching of all the data on different refinement levels into GPU memory, and depending on whether 2 or just 1 levels fit in memory, there is a varying loss in efficiency: on PCI-e cards 0.38$\times$ with 2 levels and 0.26$\times$ with 1 level, and with NVLink cards 0.6$\times$ and 0.47$\times$ respectively. They did not consider a case where not even a single level will fit in memory. Research by Buono et. al. \cite{ibmpaper} applies the streaming approach to sparse matrix-vector products, also a highly bandwidth-bound algorithm, on an NVLink system, and while they do not carry out a scalability comparison between fitting in 16 GB and not, performance is shown to be bound by NVLink bandwidth for most of the testcases.

Work by Endo \cite{stencil2} and Miki et. al. \cite{stencil1} consider a runtime and a compiler approach to out of core stencil computations on previous generations of GPUs, however the focus of their study are single-stencil applications which are very simplistic and allow arbitrary temporal blocking to improve data reuse.

\section{OPS and tiling} \label{sec/ops}
\begin{figure}[!b]\centering
\vspace{-20pt}\includegraphics[width=0.40\textwidth]{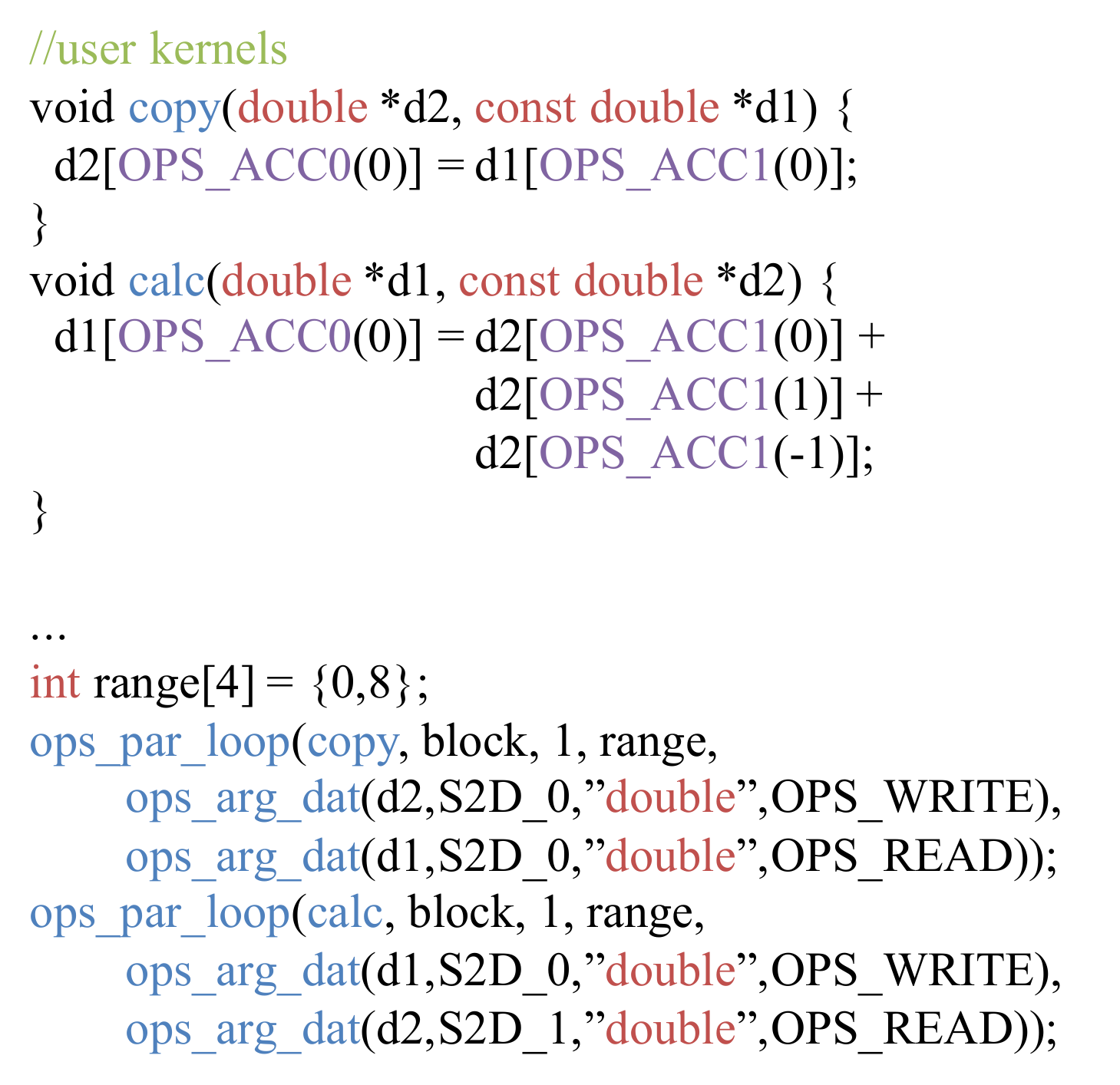}\vspace{-10pt}
\caption{\small An OPS parallel loop\normalsize}\vspace{-18pt}\label{fig/parloop}
\end{figure}

The Oxford Parallel library for Structured meshes (OPS) is a domain specific language embedded in C/Fortran, targeting multi-block structured mesh computations. Its goal is to allow domain scientists to express computations at a high level, and then facilitate data movement and parallelisation automatically at compile-time and run-time. To the user, it presents a high-level programming abstraction that can be used to describe stencil computations, and consists of the following key elements: (1) blocks, which serve to connect (2) datasets defined on these blocks, (3) stencils used to access datasets, and (4) parallel loops that iterate over a given iteration range applying a computational kernel at each point, accessing datasets defined on a given block with pre-defined stencils, also describing the type of access (read/write/both). The description of the parallel loop is perhaps the most important part of the abstraction: it means that points in the iteration space can be executed in any order, and therefore the library is free to parallelise it in any way, and to manage all data movement - a specific example is given if Figure \ref{fig/parloop}.

Initially, all data is handed to the library, and the user subsequently can only refer to them using opaque handles - returning any data to the user happens through OPS API calls, such as fetching a dataset to file, accessing a specific value, or getting the result of a reduction. The parallel loop construct contains all necessary information to execute a loop over the computational grid. This allows the library to manage all data movement and parallelisation - such as domain decomposition over MPI and halo exchanges, or explicit management of GPU memory spaces and launch of GPU kernels. It also allows OPS to delay the execution of these loops; given a sequence of loops, datasets accessed and access patterns, it is possible to carry out data dependency analysis. We use this to compute a new loop schedule that corresponds to a skewed tiling schedule, keeping all data used by a given tile resident in fast memory for the duration of that tile. The limit to the analysis of subsequent loops is any API call which returns data to the user, based on which e.g. a control decision will be made. This approach has been demonstrated to work well on large applications such as CloverLeaf and OpenSBLI \cite{arxiv} in our previous work, achieving a 2$\times$ speedup when tiling across several hundred computational loops that access tens of different datasets with tens of different stencils. We also demonstrated that stencil and polyhedral compilers such as Pluto \cite{uday08cc}, Pochoir \cite{Tang:2011:PSC:1989493.1989508}, and Halide \cite{halide} are not capable of tiling such applications, because loops are distributed across many compilation units, and the execution path cannot be determined at compile time.

\section{Tiling in stacked memory} \label{sec/adapt}

When stacked memory can be used as a last level cache, applying the tiling techniques is fairly straightforward: the tile sizes simply need to be set according to the size of the stacked memory. This is the approach we take when using Intel's Knights Landing, and as we will show it performs very well. This approach can also be used on the P100 GPU, relying on unified memory and oversubscription: whenever there is a page miss, it automatically gets transferred to the GPU memory, and gets transferred back when the CPU accesses it, or when the GPU runs out of memory - essentially making GPU memory into a large cache. As we will show this in itself is not performant enough, due to the high latency of page misses and their transfers. This approach is further complemented by prefetch commands that move memory pages in bulk between CPU and GPU with a much higher throughput.

\begin{figure}[!t]
\hspace*{-10pt}\centering
\includegraphics[width=0.35\textwidth]{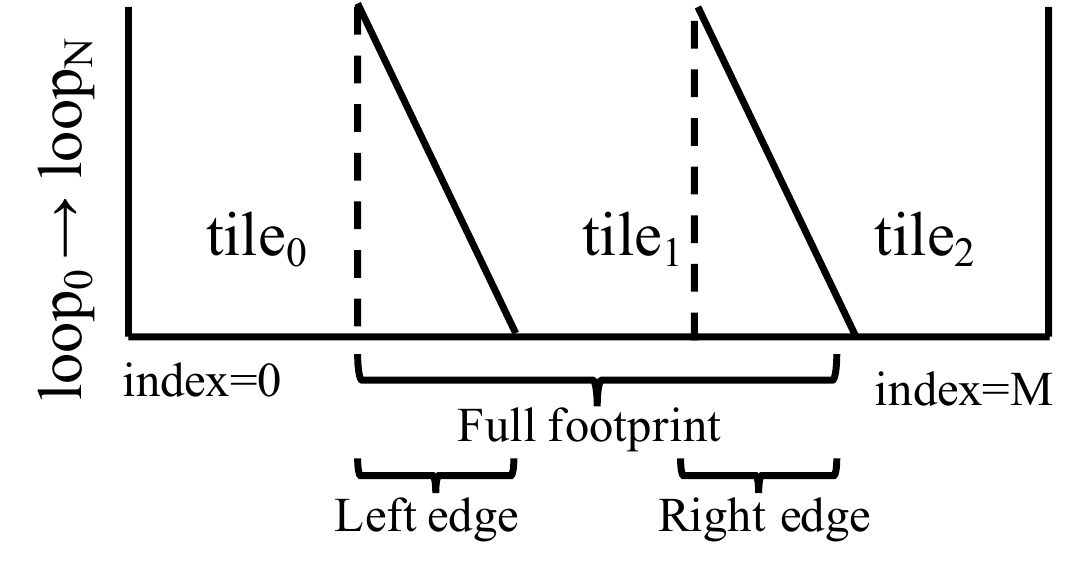}\vspace{-12pt}
\caption{\small Tile regions\normalsize}\vspace{-15pt}\label{fig/tile_layout}
\end{figure}

The alternative is to use explicit memory management, and asynchronous memory copies in particular, to move the data required for subsequent tiles back and forth. As GPUs are capable of simultaneously moving memory between CPU and GPU in both directions and running GPU kernels, we use a triple buffering (which we will call the \emph{three slots}) approach: while a given tile is executing, we are copying the results of the previous tile back to the CPU, and copying the data required by the next tile to the GPU. The overlap between tiles makes this algorithm fairly convoluted, thus we first introduce some notations, as illustrated on Figure \ref{fig/tile_layout}: the end of a given tile overlaps with the beginning of the next one - how much depends on the stencils used, we call these regions the ``left edge'' and ``right edge'' of the tiles. To denote all the data required for the execution of a given tile we use ``full footprint'', the part that omits the region that overlaps with the next tile is ``left footprint'', and the part that omits the region that overlaps with the previous tile ``right footprint'' Then the algorithm can be described in Algorithm \ref{alg}:

\begin{algorithm}
\caption{\textsc{Explicitly managed tiling algorithm}\label{alg}}
\begin{algorithmic}[1]
\FOR{$t = 0...num\_tiles$}
\STATE \COMMENT{ preparation phase }
\IF{t==0}
\STATE Upload ``full footprint'' into slot in stream 0
\ENDIF
\STATE Wait for stream 0 and 1
\STATE Upload ``right footprint'' to next slot in stream 1
\STATE Adjust base pointers of datasets for virtual position
\STATE slot++
\STATE \COMMENT{ execution phase }
\STATE Execute all loops in current tile in stream 0
\STATE \COMMENT{ finishing phase }
\STATE Wait for stream 0 and 2
\STATE For all datasets, transfer ``right edge'' of current tile to ``left edge'' of next tile in stream 0
\STATE Download ``left footprint'' of all datasets modified by current tile in stream 2
\ENDFOR
\end{algorithmic}	
\end{algorithm}

Streams 0-2 denote independent streams of operations (corresponding to CUDA streams) that can be carried out simultaneously - to satisfy data dependencies and to avoid overwriting data that is still being accessed, the appropriate synchronisations are introduced. On line 8 we have an operation that adjusts the pointers later dereferenced during execution to account for the fact that not all data is resident on the GPU. Before executing tiles (except for tile 0), the ``right edge'' of the previous tile - that is the overlapping region - needs to be copied to the ``left edge'' of the current tile, because data for each tile is kept separate to avoid any race conditions. 

\subsection{Optimisations} \label{sec:opt}

There are two basic optimisations that reduce the amount of data that needs to be moved: datasets that are read-only are not copied back to the CPU, and datasets that are written first are not uploaded to the GPU. This can significantly reduce the time taken by copies, however as we will show it can still be a bottleneck on PCI-e GPUs.

We introduce an additional optimisation, building on the fact that in most stencil codes there is a number of datasets used as temporaries within one time step, they do not carry information across time iterations - and the application itself follows a cyclic execution pattern. If the loops in the tile correspond to all the loops in the time iteration, then these datasets do not need to be copied back to the CPU, saving additional time - thus we can optionally not copy datasets back to the CPU that are written first. This of course is an unsafe optimisation, as not all stencil codes are structured like that, and even for the ones that are, datasets written in the initialisation phase of the application (e.g. calculating coordinates and volumes) will be read during the main part of the simulation. Remember, that OPS needs to execute all preceding loops when something needs to be returned to user space, therefore it cannot see ahead to determine which datasets will be read later on. In the application codes we study, we enable this optimisation by setting a flag after the initialisation phase, once the regular cyclic execution pattern begins.

Finally, we consider that when finishing the execution of a chain of loops, the processing (and even the existence) of the next chain cannot begin until the current one finished. This also means that fetching the data required for the first tile of the second loopchain will not start until the execution of the first chain's last tile finished - meaning there is no overlap in CPU-GPU copies for the first tile and GPU execution of the last tile. To address this, we implement a speculative prefetching scheme: during the execution of the first chain's last tile, we start uploading data for the second chain's first tile; but of course without information on what the second loopchain looks like - therefore we use the first tile's dependencies, assuming that the second chain of loops will look similar to the first. Then, when the processing of the second chain of loops actually starts, we check what was uploaded previously, and upload anything that is missing.

\section{Performance analysis} \label{sec/perf}

In this section we first introduce the stencil applications being studied, then move on to analyse performance and size scalability on the KNL, then on the P100 GPU using various data movement strategies.

\subsection{Stencil codes}

Our first two applications benchmarked are the 2D and the 3D versions of CloverLeaf \cite{cloverleaf}, mini-applications from the Mantevo suite \cite{mantevo} that solve the compressible Euler equations on a Cartesian grid, using an explicit second-order method. CloverLeaf uses an explicit time-marching scheme, computing energy, density, pressure and velocity on a staggered grid, using the finite volume discretisation. One timestep involves two main computational stages: a Lagrangian step with a predictor-corrector scheme, advancing time, and an advection step - with separate sweeps in the horizontal/vertical/depth dimensions. The full source of the original is available at \cite{cloverleaf-github}.

CloverLeaf 2D/3D has 25/30 variables per gridpoint (the number of datasets), and there are 30/46 multi-point stencils used to access them at different stages of the computation. There are a total of 83/141 parallel loops over the grid, and these are spread across 15 different source files - there is significant branching between loops, depending on e.g. sweep direction. A single time iteration consists of a chain of 153/603 parallel loops to be performed in sequence.

The third stencil application is OpenSBLI \cite{opensbli}, a large-scale academic research code being developed at the University of Southampton, focusing on the solution of the compressible Navier-Stokes equations with application to shock-boundary layer interactions (SBLI). Here we are evaluating a 3D Taylor-Green vortex testcase, which consists of 27 nested loops over the computational grid, using 9 different stencils and accessing 29 datasets defined on the 3D grid. This code uses a third-order Runge-Kutta scheme without adaptive step control, and does not use any reductions during the bulk of the computations, therefore we can practically tile across an arbitrary number of loops - this will be explored during performance analysis.

Considering all of these applications are bound my memory bandwidth, due to a low flop/byte ratio, the key performance metric is achieved bandwidth - this is what we report in this paper. Bandwidth is calculated by looking at the iteration range of each loop, and the datasets it accesses, thus calculating the number of bytes moved (1$\times$ multiplier for reads or writes, and 2$\times$ for reads and writes). This is then divided by the runtime of the loop to get GB/s. This is finally weighted averaged over all loops for the entire application to produce the ``Average Bandwidth'' metric that we report.

These applications are representative of structured mesh stencil computations particularly in the Computational Fluid Dynamics area: large numbers of variables per gridpoint, many different sweeps over the computational grid, and a low arithmetic intensity. They however may not be of representative of different classes of applications (e.g. Lattice-Boltzmann) that have different characteristics.

\subsection{Tiling on the Knights Landing}

On Intel's KNL architecture, it is possible to use the integrated MCDRAM as a separate memory space (Flat mode), by either using different \emph{malloc} calls, or by using a tool like \emph{numactl} to force all allocations to it. Using the same tools, one can also not use the MCDRAM at all, so all memory allocation and movement goes to DDR4. MCDRAM can also be used as a further level of cache between L2 (no L3 cache on the KNL) and DDR4. In this work, we use the MCDRAM in the quadrant clustering mode, which affects memory accesses between different cores (for details, see \cite{cachemodes}), as tests have shown it to perform better on these applications than any of the other settings.

To explore performance and to demonstrate the relevance and benefits of our work, we run our benchmarks at different sizes in four configurations: in flat mode only using DDR4 without tiling, in flat mode only using MCDRAM without tiling, in cache mode without tiling, and in cache mode with tiling enabled. We did not realise any performance benefit from trying to tile in L2 cache because of the large data footprint of our tiles and the relatively small amount of L2 cache per core, therefore we do not discuss that option. The best performance is expected from the flat MCDRAM configuration, and the worst from the flat DDR4 configuration, in caching mode -- with or without tiling, the performance should be between these two. We use 4 MPI processes, with 32 threads each, pinned to cores in different quadrants, on an Intel Xeon Phi x200 7210 processor, running CentOS 7. STREAM Triad bandwidth on this machine is 291 GB/s in cache mode, and in flat mode when \emph{malloc} is used for memory allocation, DDR4 bandwidth is 60.8 GB/s and MCDRAM bandwidth is 314 GB/s. Please note that this is a modified version of the STREAM benchmark - the original achieves up to 450 GB/s in MCDRAM, however it allocates memory statically, which is  unrealistic actual applications which allocate memory dynamically.

\begin{figure}[!t]
\hspace*{-10pt}\centering
\includegraphics[width=0.45\textwidth]{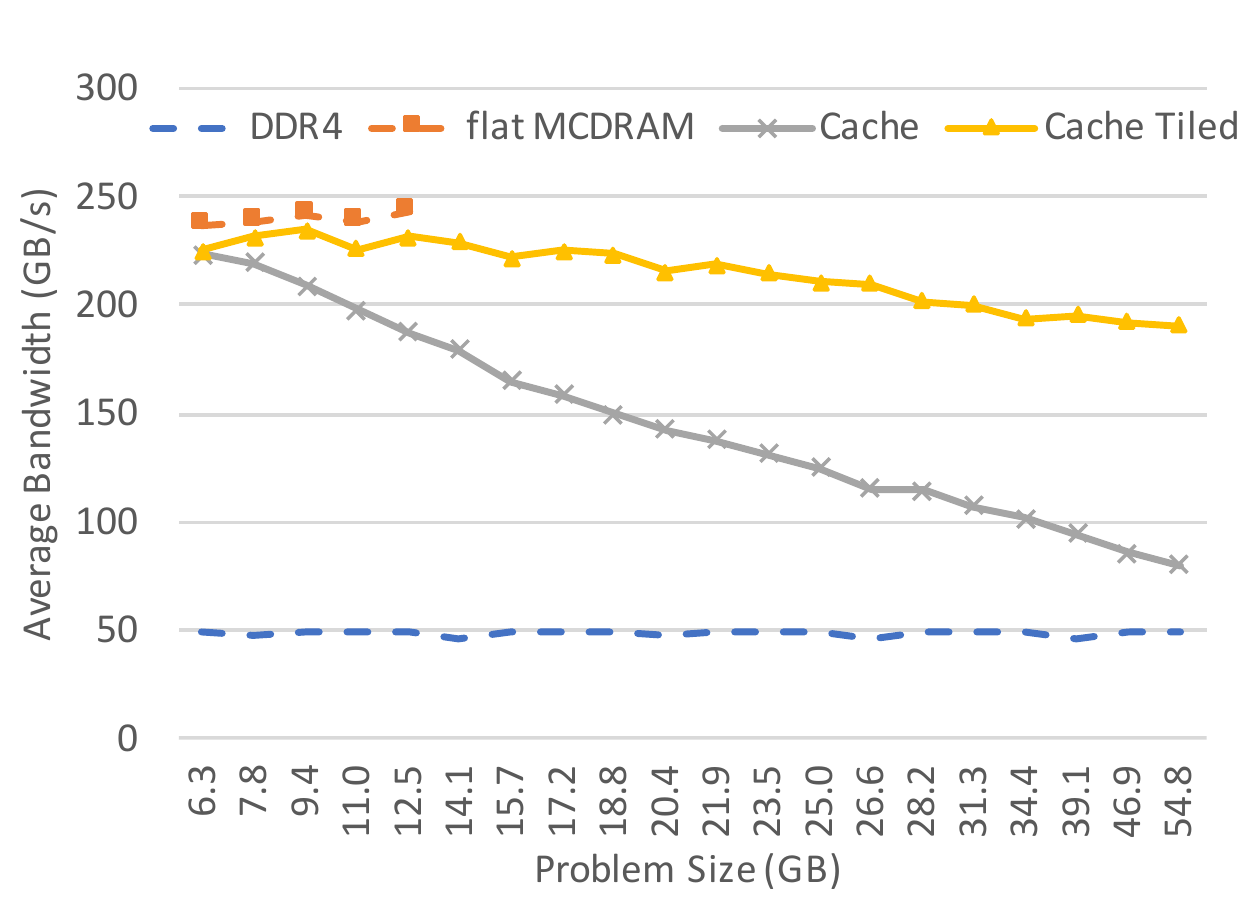}\vspace{-12pt}
\caption{\small CloverLeaf 2D problem scaling on the KNL\normalsize}\vspace{-5pt}\label{fig/knl_c2d}
\end{figure}

\begin{figure}[!t]
\hspace*{-10pt}\centering
\includegraphics[width=0.45\textwidth]{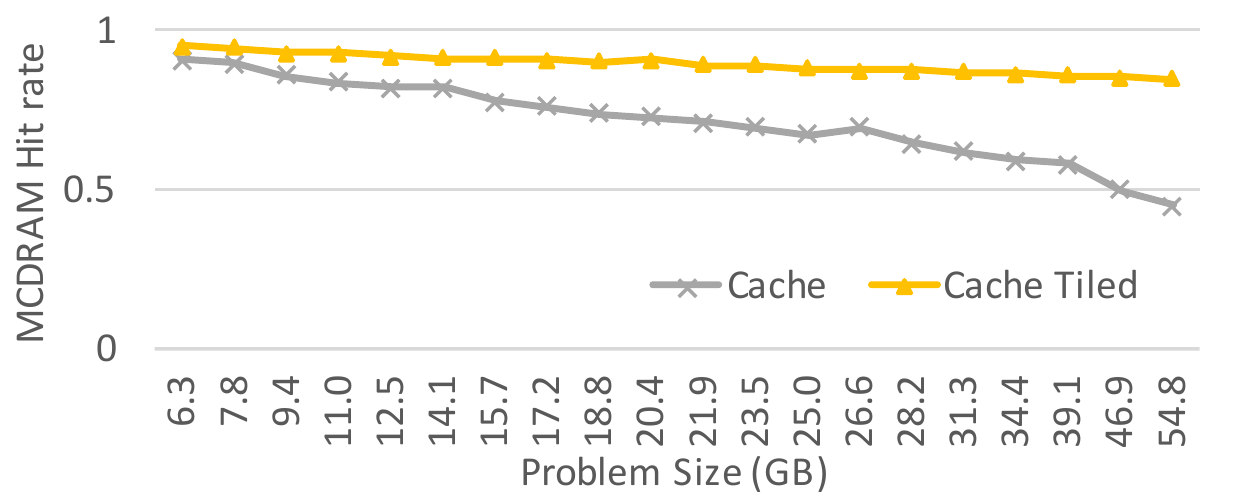}\vspace{-12pt}
\caption{\small MCDRAM Cache Hit rate on CloverLeaf 2D as reported by PCM\normalsize}\vspace{-5pt}\label{fig/knl_c2d_hitrate}
\end{figure}

\begin{figure}[!t]
\hspace*{-10pt}\centering
\includegraphics[width=0.45\textwidth]{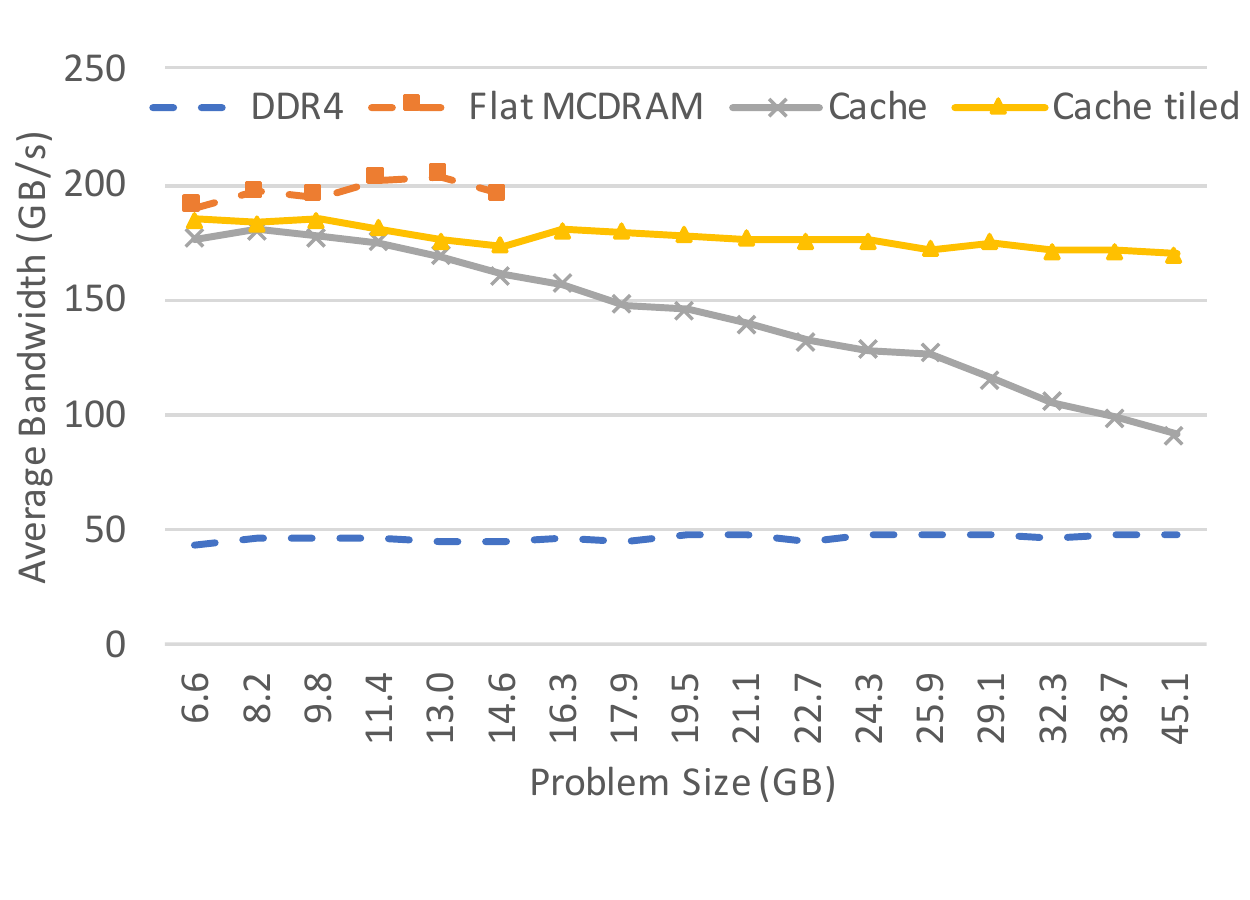}\vspace{-25pt}
\caption{\small CloverLeaf 3D problem scaling on the KNL\normalsize}\vspace{-12pt}\label{fig/knl_c3d}
\end{figure}
\begin{figure}[!t]
\hspace*{-10pt}\centering
\includegraphics[width=0.45\textwidth]{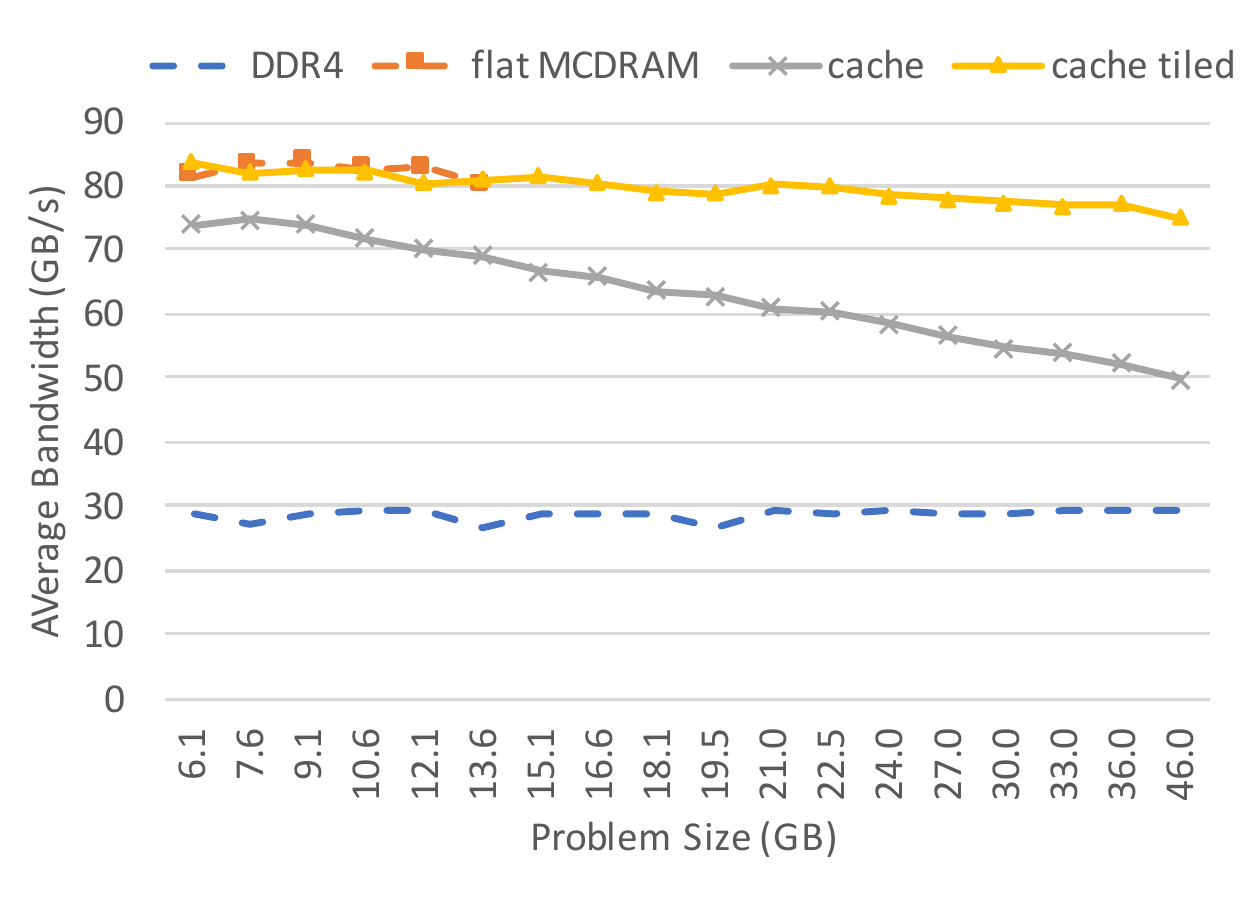}\vspace{-12pt}
\caption{\small OpenSBLI problem scaling on the KNL\normalsize}\vspace{-15pt}\label{fig/knl_rs}
\end{figure}

Figures \ref{fig/knl_c2d}, \ref{fig/knl_c3d}, and \ref{fig/knl_rs} show the average bandwidth over the entirety of the application - the same trends appears on all three applications: in the flat configurations, performance holds steady, but of course for MCDRAM we quickly run out of available memory, and trying to run larger problems leads to segmentation faults. On CloverLeaf 2D and 3D using only DDR4 shows an average of 50 GB/s, but in the flat MCDRAM configuration there is a 20\% difference between 2D and 3D (200 vs 240 GB/s) - this is due to the 3D version having more complex kernels that are more sensitive to latency - MCDRAM is 4.8$\times$ (2D) and 4$\times$ (3D) faster than DDR4 . This is even more pronounced for OpenSBLI, where a single large kernel, very sensitive to latency, accounts for 60\% of the runtime: with DDR4 30 GB/s and with MCDRAM 83 GB/s is achieved.

Switching MCDRAM to cache mode and running increasingly larger problems shows a graceful degradation of performance: at small problem sizes there is very little drawback of using cache mode instead of flat mode, and scaling to a size of 48 GB, or 3 times larger than the cache, shows slowdowns of 0.36$\times$ (86 GB/s) for CloverLeaf 2D, 0.45$\times$ (98 GB/s) for CloverLeaf 3D, and 0.59$\times$ (50 GB/s) for OpenSBLI. The hit rates in MCDRAM cache for CloverLeaf 2D are shown in Figure \ref{fig/knl_c2d_hitrate}; they show a steady decline matching runtimes.

\begin{figure*}
\hspace*{-10pt}\centering
\includegraphics[width=0.9\textwidth]{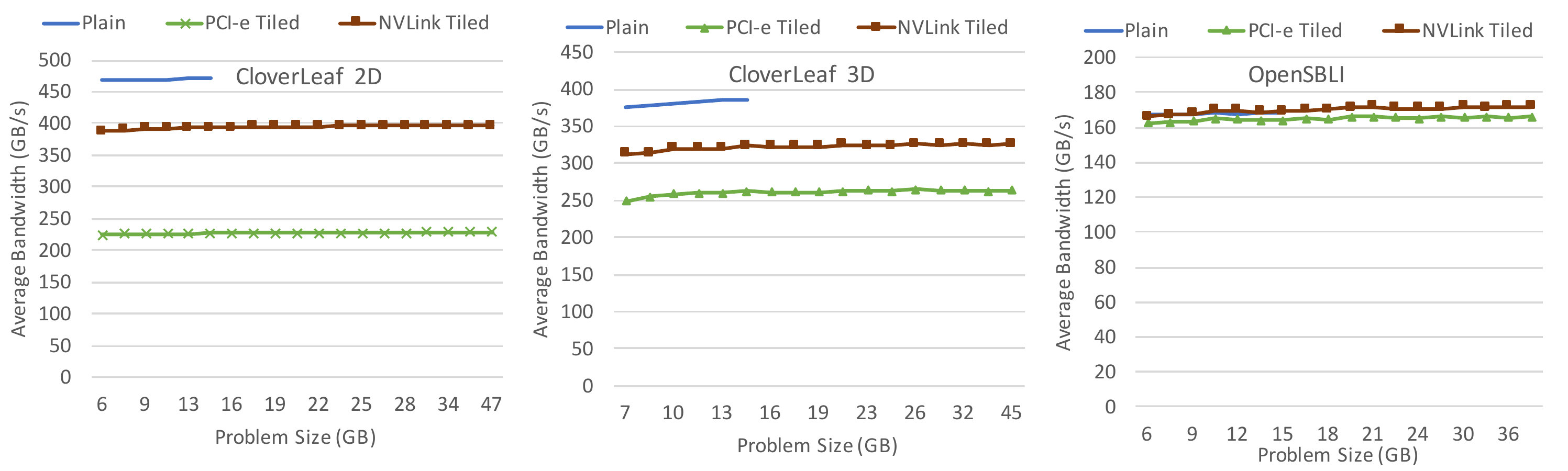}\vspace{-12pt}
\caption{\small Problem scaling on the P100 GPU\normalsize}\vspace{-15pt}\label{fig/gpu_all}
\end{figure*}

Enabling tiling has two key effects on the KNL: improving locality in MCDRAM cache, and improving MPI communications. The latter is because without tiling, OPS will exchange the halos of datasets on a per-loop basis, whereas with tiling it calculates the halos to be exchanged needed for the execution of the entire loop chain - thus halo exchanges happen once at the beginning of each loop chain; they are larger in size than any individual exchange without tiling, but much fewer in number.  As we are running with 4 MPI processes the latter is important, and it accounts for the performance difference at small problem sizes that would otherwise fit in the 16 GB cache. With tiling there is very little performance loss at problem sizes far exceeding the 16 GB cache: comparing the smallest problems (6GB) to the 48 GB problems, there is only a 15\% decrease in efficiency on CloverLeaf 2D, 7\% on CloverLeaf 3D, and 7\% on OpenSBLI. In total, our tiling algorithms have improved performance at the 48 GB size by 2.2$\times$ on CloverLeaf 2D, 1.7$\times$ on CloverLeaf 3D, and 1.5$\times$ on OpenSBLI compared to the non-tiled version. Hit rates in MCDRAM cache are shown in Figure \ref{fig/knl_c2d_hitrate} for CloverLeaf 2D; they show a slow decrease matching that of runtime.

\subsection{Tiling on GPUs with Explicit Memory Management}

To evaluate performance on GPUs, we use NVIDIA Tesla P100 cards, one with PCI-e, connected to a single-socket Xeon E5-1660 v4, running Ubuntu 16.04. The other P100 card is connected via NVLink to a Power8 CPU (IBM Minsky system), running an Ubuntu 16.04 system. For both, we use CUDA 8, driver version 375.39. The device-to-device streaming copy bandwidth measured is 509.7 GB/s.

We have developed several ways of running larger problems than 16GB on GPUs, relying on either unified memory, or explicit memory management. In Section \ref{sec:opt} we presented a number of optimisations that help improve performance when using explicit memory management: (1) less movement of read-only and write-first data, (2) discarding the values of write-first data (i.e. temporary datasets) during cyclic execution (Cyclic/NoCyclic), and (3) speculative prefetching of data for the next chain of loops (Prefetch/NoPrefetch). In this section we will explore the implications of these approaches, except for (1) which is enabled all the time.


Overall best performance with and without tiling are shown in Figure \ref{fig/gpu_all}. As on the KNL platform, we see some differences in baseline performance (without tiling, but only up to 16GB) between the various applications: CloverLeaf 2D achieves an average of 470 GB/s, CloverLeaf 3D achieves only 380 GB/s, due to its more complex computations, and OpenSBLI achieves 170 GB/s, due to 68\% of the runtime being spent in a particularly computationally intensive and latency sensitive kernel (the average bandwidth of all the other kernels is 450 GB/s). 


When tiling is enabled with all the optimisations, we can see that there is still a gap in performance for CloverLeaf between the baseline and the tiled versions, but not so for OpenSBLI. This ultimately comes down to whether there are enough computations in the loop chains being tiled over to hide the cost of moving data. With OpenSBLI we can tile over an arbitrary number of loops - here we do so over 3 time iterations (each with 3 Runge-Kutta steps), and so memory movement can be completely hidden (NVLink performance is slightly higher due to higher graphics clock speeds). Baseline performance up to 16 GB matches NVLink performance.

On CloverLeaf 2D, the NVLink card achieves 84\% of the performance of the baseline, but the PCI-e card achieves only 48\%. The difference between the cards is simply due to transfer speed - while NVLink throughput averages at 30 GB/s, PCI-e throughput is only 11 GB/s. The gap of 16\% between tiling on the NVLink card and the baseline is due the fact that every 10 iterations the application calculates a number of variables summarising the computational field, such as pressure, kinetic energy, etc., resulting in a one-long loop chain reading a large number of datasets with a very poor copy/compute overlap. 

On CloverLeaf 3D, the NVLink card achieves the same 84\% of the performance of the baseline, but the PCI-e card achieves a higher 68\%. The differences come down to the same reasons, but with the 3D application, there is much more data reuse (thanks to a larger number of loops), therefore the PCI-e card achieves a much higher efficiency.

\begin{figure}
\hspace*{-10pt}\centering
\includegraphics[width=0.40\textwidth]{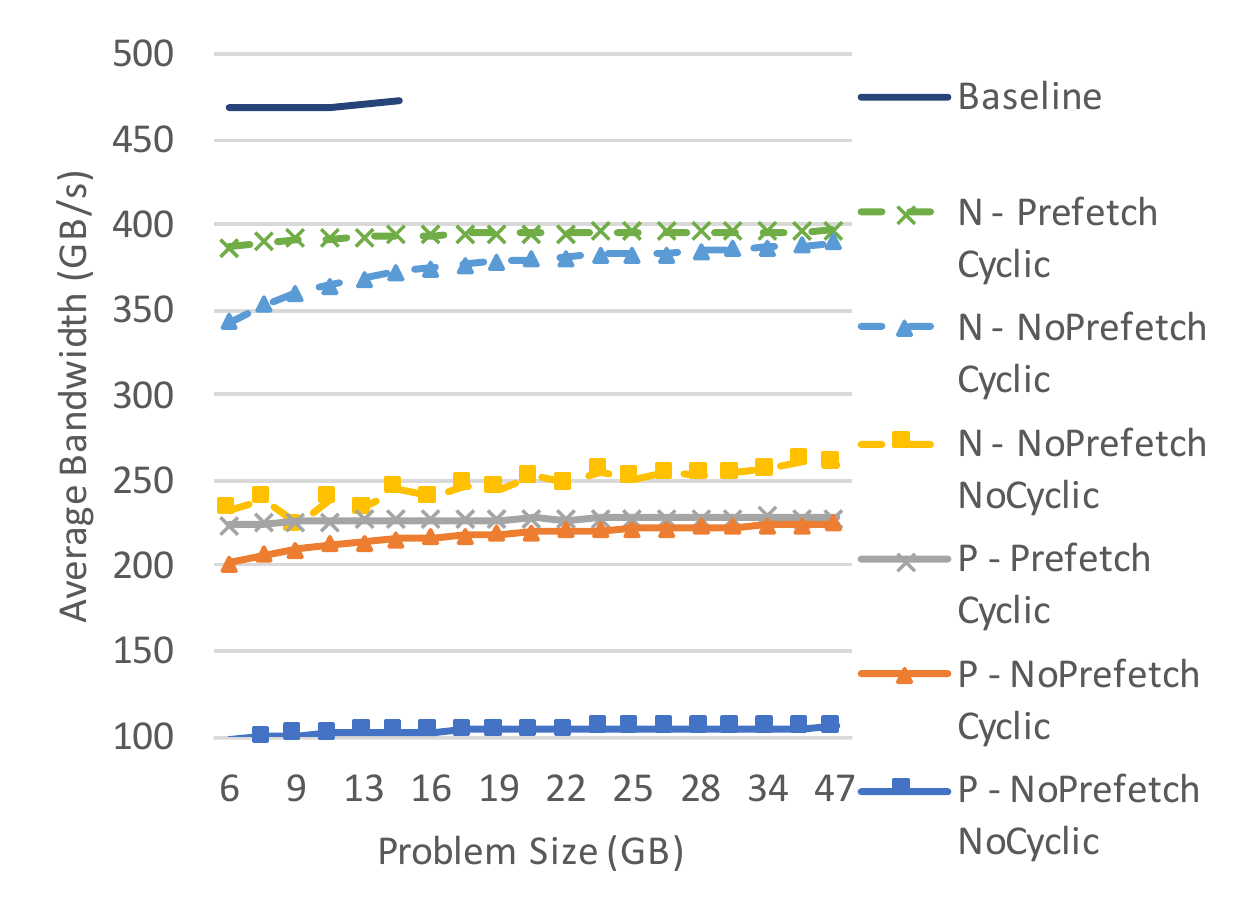}\vspace{-12pt}
\caption{\small Tiling optimisations on CloverLeaf 2D on the P100 (P-PCI-e, N-NVLink)\normalsize}\vspace{-15pt}\label{fig/gpu_opt_c2d}
\end{figure}
\begin{figure}
\hspace*{-10pt}\centering
\includegraphics[width=0.40\textwidth]{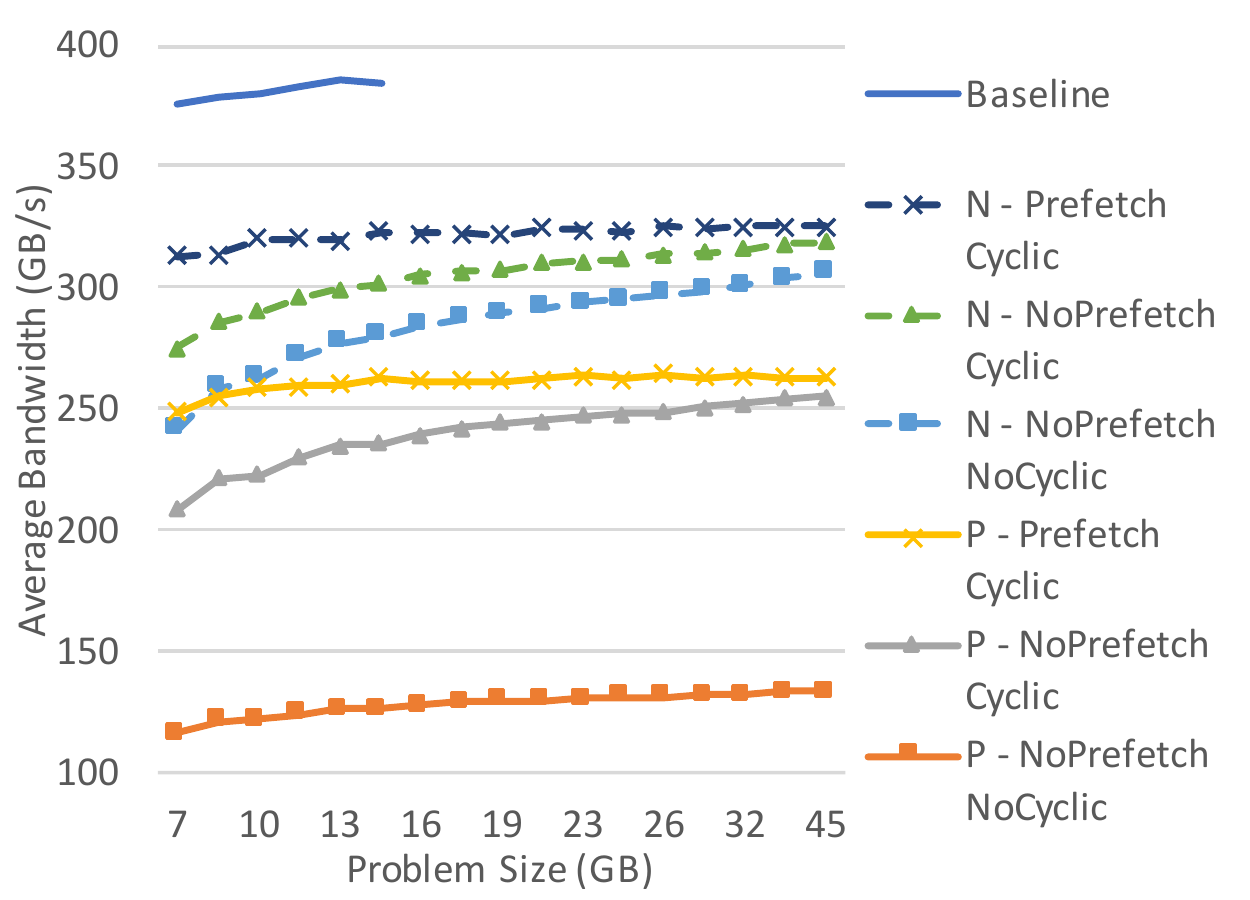}\vspace{-12pt}
\caption{\small Tiling optimisations on CloverLeaf 3D on the P100 (P-PCI-e, N-NVLink)\normalsize}\vspace{-15pt}\label{fig/gpu_opt_c3d}
\end{figure}
\begin{figure*}
\hspace*{-10pt}\centering
\includegraphics[width=0.85\textwidth]{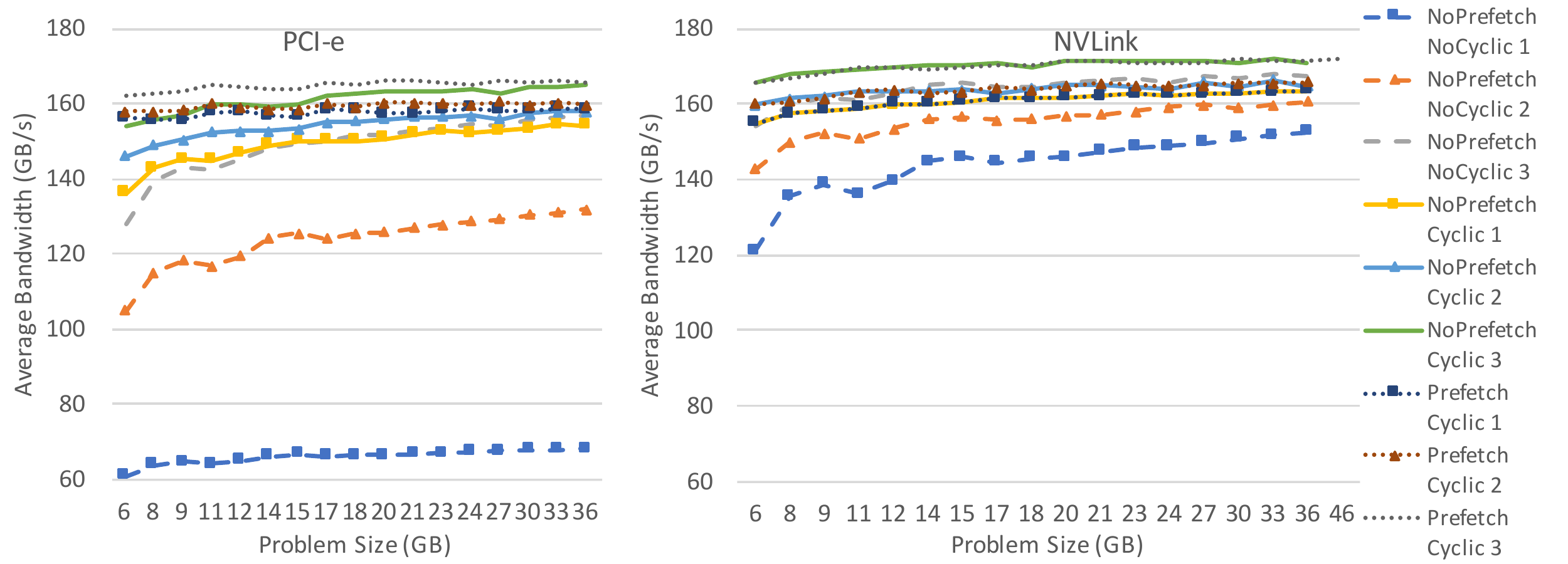}\vspace{-12pt}
\caption{\small Tiling optimisations on OpenSBLI on the P100\normalsize}\vspace{-10pt}\label{fig/gpu_opt_rs}
\end{figure*}

Delving into the optimisations, we show results on three combinations: without the speculative prefetch of tile 0, and without/with the assumption of cyclic behaviour and skipping the download of write-first data (NoPrefetch NoCyclic/Cyclic), and third, with both enabled (Prefetch Cyclic). Performance figures on CloverLeaf 2D/3D shown in Figures \ref{fig/gpu_opt_c2d} and \ref{fig/gpu_opt_c3d} clearly show the importance of reducing memory movement through the Cyclic optimisation, particularly for the 2D application where data reuse within a tile is less than for the 3D application. For 3D, on the NVLink card, the benefit is smaller, due to the interconnect being quite fast already. Enabling the prefetching optimisation as well is particularly beneficial at smaller problem sizes: prefetching helps hide the latency of moving memory for the first tile - as there are many more tiles at larger problem sizes, this latency is proportionally smaller.

For OpenSBLI, we can control how many loops to tile over: we experiment with tiling over 1, 2, or 3 timesteps. Enabling the Cyclic and Prefetch optimisations have the same effects as on the CloverLeaf codes: more pronounced on the PCI-e card, and particularly important at smaller problem sizes. By tiling over more loops, we can improve the reuse of data within the tiles, and allow for more time to hide the latency of data movement between CPU and GPU.

\subsection{Tiling with Unified Memory}\label{sec/um}

Unified Memory, introduced in CUDA 6, greatly simplifies memory management for GPUs: memory accessed on the GPU will be automatically transferred to the GPU if it's not there yet. Prior to the Pascal generation of GPUs, one was not allowed to oversubscribe GPU memory by allocating more ``Managed'' memory on the CPU than the memory size of the GPU. When the full problem size is less then the size of GPU memory, there is some initial overhead in transferring data to the GPU, but afterwards all the data will stay on the GPU (unless accessed on the CPU), therefore performance can be expected to be the same as with explicitly managed memory.

\begin{figure*}
\hspace*{-10pt}\centering
\includegraphics[width=0.9\textwidth]{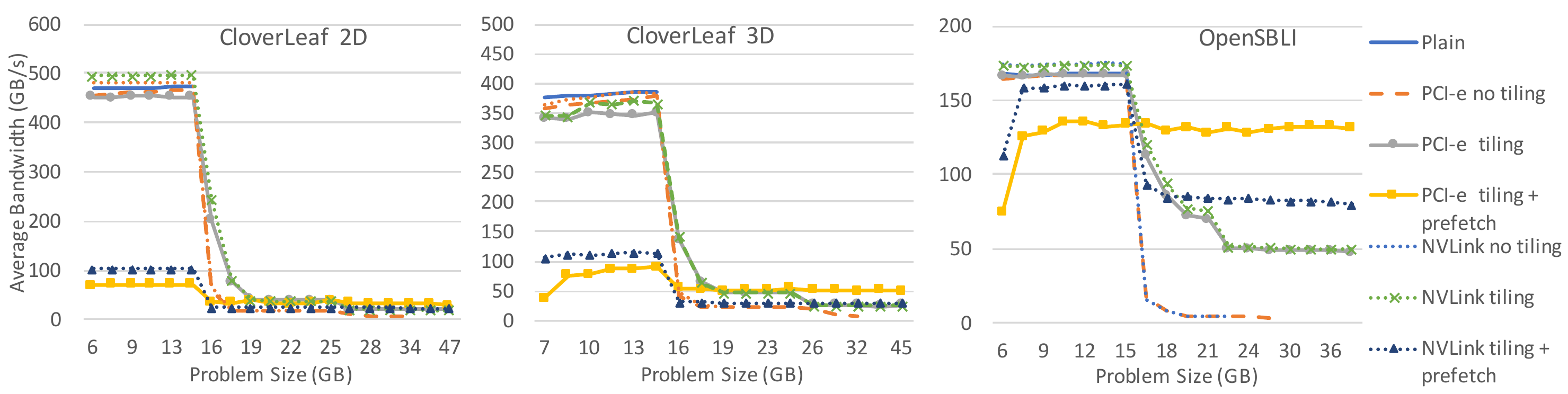}\vspace{-12pt}
\caption{\small Problem scaling with Unified Memory on the P100 GPU\normalsize}\vspace{-20pt}\label{fig/unified}
\end{figure*}

First, we evaluate performance relying only on automatic page migration, with and without tiling. As Figure \ref{fig/unified} shows, performance matches the baseline up to 16GB, after which there is a dramatic drop in performance. This is due to the high latency of page migration - the throughput is the same on both NVLink and PCI-e devices, suggesting that data movement due to page misses are bound by latency not bandwidth.

The tiled version does perform up to 3$\times$ better than the non-tiled version, absolute performance is still poor. This can be further improved by the use of \texttt{cudaMemPrefetch} prefetch commands, which instruct the driver to migrate pages to or from the GPU. Unlike regular \texttt{cudaMemcpy} they involve considerable CPU work within the GPU driver, which also means that if not used carefully, they will not overlap with one another, nor with kernel executions. We achieved best performance when prefetches to the host were issued right after all computational kernels for a given tile were queued, in the same stream, and prefetches to the device were issued at the same time in a different stream, followed by a synchronisation on that stream - leading to the first set to be executed by the deferred pathway in the driver, and the second set by the non-deferred pathway. After each tile, the streams are swapped, so the next tile starts execution in a third, so far idle, stream. Unfortunately the performance of prefetches drops significantly once we start oversubscribing memory, which is another issue in the current drivers. As Figure \ref{fig/unified} shows, the prefetch version is significantly faster above 16GB, but on CloverLeaf there is simply not enough data re-use to hide the latency of memory movement. On OpenSBLI, tiling over 5 time iterations, there is enough data re-use, but due to overlap issues  between loop chains (going from last tile of previous chain to first tile of the next) performance still does not reach that of the baseline.

Unfortunately the issues with page migration limit the usability of this approach - the problems are even worse on the IBM+NVLink platform: while below 16 GB it is faster than PCI-e, when oversubscribing memory it performs consistently worse.


\section{Conclusion} \label{sec/conclusions}
In this paper, we have presented algorithmic techniques and optimisations that allow the application of a cache-blocking tiling technique to large scale stencil codes running on architectures with small but high-bandwidth memory. We propose to use the high-bandwidth memory as a last level cache, forming large tiles across a number of loop nests in case of problems that would otherwise not fit in this memory.

We developed algorithms to explicitly manage the memory on GPUs, streaming in and out data required by different tiles, and introduce optimisations to help reduce the amount of data moved: not moving read-only data back to the CPU, not copying write-first data to the GPU, speculative prefetching of data for subsequent loop chains, and not copying temporary data back to the CPU.

After implementing the proposed algorithms into the OPS domain specific language, we carried out a detailed study of three large stencil codes: CloverLeaf 2D, CloverLeaf 3D, and OpenSBLI. Running on Intel's Knights Landing, we demonstrate how at increasing problem sizes performance drops without tiling, and that with tiling efficiency can be maintained with very little loss: at a 48GB size, 16\% loss compared to 6GB on CloverLeaf 2D, and 7\% on CloverLeaf 3D and OpenSBLI. On NVIDIA's P100 GPUs connected to the CPU via either PCI-e or NVLink, we evaluate performance using explicit memory management, as well as unified memory. Due to a number of inefficiencies and issues in the driver and the handling of prefetches, the unified memory versions do not perform well, but the explicit memory management versions do get performance close to what is achieved on small problems: on NVLink cards within 16\% on CloverLeaf 2D and 3D, and matching performance on OpenSBLI. This essentially comes down to data re-use and computational intensity in loop chains: while there is enough in OpenSBLI, there are some loop chains in CloverLeaf with low data re-use.

Our results demonstrate that it is possible to run much larger problems on architectures with high bandwidth memory than what can fit in this memory, at only a minor loss in efficiency - even in case of a class of problems which is limited by bandwidth. This work also underlines the utility of domain specific languages: to achieve these results, we did not have to modify the high-level scientific code, only components of the OPS library. Next, we would like to explore scaling to hundreds or thousands of KNLs and GPUs, and further improving MPI communications with latency hiding by computing tiles that do not depend on halo data first.

\vspace{-5pt}
\section*{Acknowledgments} \vspace{-5pt}

The authors would like to thank IBM (J\'ozsef Sur\'anyi, Michal Iwanski) for access to a Minsky system, as well as Nikolay Sakharnykh at NVIDIA for the help with unified memory performance. This paper was supported by the J\'anos B\'olyai Research Scholarship of the Hungarian Academy of Sciences. The authors would like to acknowledge the use of the University of Oxford Advanced Research Computing (ARC) facility in carrying out this work. \url{http://dx.doi.org/10.5281/zenodo.22558}. The OPS project is funded by the UK Engineering and Physical Sciences Research Council projects EP/K038494/1, EP/K038486/1, EP/K038451/1 and EP/K038567/1 on ``Future-proof massively-parallel execution of multi-block applications'' project.

\bibliographystyle{ACM-Reference-Format}
\vspace{-10pt}\bibliography{tiling}

\end{document}